\documentclass[preprint2]{aastex}

\shortauthors{R.~O.~Gray et al.}
\shorttitle{NStars Project}

\begin{document}

\title{Contributions to the Nearby Stars (NStars) Project: Spectroscopy
of Stars Earlier than M0 within 40 parsecs: The Northern Sample I.}

\author{R.O. Gray}
\affil{Department of Physics and Astronomy}
\affil{Appalachian State University, Boone, NC 28608}
\email{grayro@appstate.edu}
\author{C.J. Corbally}
\affil{Vatican Observatory Research Group, Steward Observatory}
\affil{Tucson, AZ 85721-0065}
\email{corbally@as.arizona.edu}
\author{R.F. Garrison}
\affil{David Dunlap Observatory, Richmond Hill, Ontario}
\email{garrison@astro.utoronto.ca}
\author{M.T. McFadden}
\affil{Department of Physics and Astronomy}
\affil{Appalachian State University, Boone, NC 28608}
\email{mcfaddnm@pm.appstate.edu}
\author{P.E. Robinson}
\affil{Department of Physics and Astronomy}
\affil{Appalachian State University, Boone, NC 28608}
\email{}

\begin{abstract}

We have embarked on a project, under the aegis of the Nearby Stars (NStars)/
Space Interferometry Mission Preparatory Science  
Program to obtain spectra, spectral types, and, where feasible, basic physical 
parameters for the 3600 dwarf and giant stars earlier than M0 within 40
parsecs of the sun.  In this paper we report on the results of this project
for the first 664 stars in the northern hemisphere.  These results include 
precise, homogeneous spectral types, basic physical parameters (including 
the effective temperature, surface gravity and the overall metallicity, 
[M/H]) and measures of the chromospheric activity of our program stars.  
Observed and derived data presented in this paper are also available on the 
project's website http://stellar.phys.appstate.edu/.

\end{abstract}


\section{Introduction}

The three institutions represented by the authorship of this paper are
cooperating on a project under the NASA/JPL Nearby Stars / Space Interferometry
Mission Preparatory Science program to obtain spectroscopic observations
of all 3600 main-sequence and giant stars with spectral types earlier than
M0 within a radius of 40 pc.  We are obtaining blue-violet spectra at 
classification resolution (1.5 -- 3.6\AA) for all of these stars.  These
spectra are being used to obtain homogeneous, precise, MK spectral types. In 
addition, these spectra are being used in conjunction with synthetic spectra
and existing intermediate-band Str\"omgren
$uvby$ and broad-band $VRI$ photometry to derive the basic astrophysical 
parameters (the effective temperature, gravity and overall metal abundance [M/H]) for many of these stars.  We are 
also using these spectra, which include the \ion{Ca}{2} K and
H lines, to obtain measures of the chromospheric activity of the program stars
on the Mount Wilson system.  The purpose of this project is to provide data
which will permit an efficient choice of targets for both the Space
Interferometry Mission (SIM) and the projected Terrestrial Planet Finder
(TPF).  In addition, combination of these new data with kinematical data
should enable the identification and characterization of stellar 
subpopulations within the solar neighborhood.

Observations for this project are being carried out on the 1.9-m telescope
of the David Dunlap Observatory in a northern polar cap (${\rm DEC}  > 
+50^\circ$), the 0.8-m telescope of the Dark Sky Observatory 
($-10^\circ \le {\rm DEC} \le +50^\circ$), the 2.3-m Bok telescope of
Steward Observatory ($-30^\circ \le {\rm DEC} \le -10^\circ$) and
the 1.5-m telescope at Cerro Tololo Interamerican Observatory (${\rm DEC}
\le -30^\circ$), although there is considerable overlap between all
of these samples.  In this paper, we report on results for the first
664 stars, all observed at the Dark Sky Observatory. 

\section{Observations and Calibration}

The observations reported in this paper were all made on the 0.8m telescope 
of the Dark Sky Observatory (Appalachian State University) situated on
the escarpment of the Blue Ridge Mountains in northwestern North Carolina.  
The Gray/Miller
classification spectrograph was employed with two gratings with 
600 grooves mm$^{-1}$ and 1200 grooves mm$^{-1}$ and a thinned, 
back-illuminated 1024 $\times$ 1024 Tektronix CCD operating in the
multipinned-phase mode.  The observations were made with a 100$\mu$m slit,
which corresponds to approximately 2$^{\prime\prime}$ at the focus of the
telescope; average seeing at the Dark Sky Observatory (DSO) is about 
3$^{\prime\prime}$.  The 100$\mu$m slit with the two gratings yields 
2-pixel resolutions of
3.6\AA\ and 1.8\AA\ and spectral ranges of 3800 -- 5600\AA\ and
3800 -- 4600\AA\ respectively.  The lower-resolution spectra were used 
primarily for the late-type stars in the sample (later than G5) whereas the
higher-resolution spectra were used primarily for the earlier-type stars.  An
iron-argon hollow-cathode comparison lamp was used for the wavelength 
calibration, and all the spectra were reduced with standard methods using 
IRAF\footnote{IRAF is distributed by
the National Optical Astronomy Observatories (NOAO).  NOAO is operated by 
the Association of Universities for Research in Astronomy (AURA), Inc. 
under cooperative agreement with the National Science Foundation}.

The 1.8\AA\ resolution spectra were rectified using an X-windows program,
{\tt xmk19}, written by one of us (ROG) and were used in that format for both
spectral classification and for the determination of the basic physical
parameters.  For the late-type stars rectification is problematical as
no useful ``continuum'' points can be identified.  In addition, the energy 
distribution contains useful information for both the spectral
classification and the determination of the basic physical parameters.  We have
therefore made an attempt to approximately flux calibrate the 3.6\AA\
resolution spectra even
though they were obtained with a narrow slit.  During the course of our
observations for this project at DSO over the past 
three years, we have, at intervals of a few
months, made observations of spectrophotometric standards at a variety
of airmasses.  These standard observations have been used to approximately
remove the effects of atmospheric extinction and to calibrate the
spectrograph throughput as a function of wavelength.  Except for observations
made at high airmasses ($> 1.8$ airmasses) this procedure yields
calibrations of {\it relative} fluxes with accuracies on the order of 
$\pm 10$\%.  While this
is sufficient for the purposes of accurate spectral classification, the 
determination
of the basic physical parameters requires a more accurate flux calibration.
This we have achieved by ``photometrically correcting'' the fluxes using
Str\"omgren photometry.  The DSO 3.6\AA\ resolution spectra 
have a spectral range which includes three of the four Str\"omgren bands, 
$v$, $b$ and $y$.  We perform numerical photometry on these
spectra and use the absolute flux calibration of the Str\"omgren system
\citep{gray98} to derive flux corrections at the effective wavelengths of the
three photometric bands.  Interpolation and extrapolation of these corrections
yield flux corrections over the entire observed spectrum.  From comparison
with spectrophotometric observations in the literature, we find that this 
procedure yields not only relative but also absolute fluxes with accuracies 
of about $\pm 3$\% over nearly the entire spectral range.  Fortunately, 
most of 
our program stars with spectral types of K3 and earlier have Str\"omgren 
photometry.  For the later-type stars without Str\"omgren photometry, other 
considerations (see \S4) preclude the derivation of basic physical 
parameters from our spectra and so the lack of accurate fluxes is not 
otherwise limiting.  

All of the spectra obtained for this project are available on the project's
website\footnote{http://stellar.phys.appstate.edu/}.  Rectified spectra from
DSO have an extension of {\tt .r18} or {\tt .r36} depending on the
resolution.  The 3.6\AA\ resolution flux spectra, which have not been
photometrically corrected, are normalized at a common point (4503\AA) and
have an extension of {\tt .nor}, whereas photometrically corrected spectra
are available in a normalized format ({\tt .nfx}) and in terms of absolute
fluxes ({\tt .flx}) in units of erg s$^{-1}$ cm$^{-2}$ \AA$^{-1}$.  Spectra
obtained at the other observatories (see \S1) are also available
on this website, and will be the subject of future papers.

\section{Spectral Classification}

The spectral types for the stars in this paper were obtained using
MK standard stars selected from the list of ``Anchor Points of the
MK System'' \citep{garrison94}, the Perkins catalog \citep{keenan89} and, for
late K and early M-type dwarfs from \citet{henry02}.  A list of the standards
used (and the actual spectra) can be found on the project website.

The program stars were first classified independently on the computer screen
by eye (using the graphics program {\tt xmk19}) by at least two
of the authors and then the spectral types were compared and iterated
until complete agreement was obtained.  There are significant overlaps
between the samples observed with the four telescopes employed for this
project to ensure homogeneity of our spectral types over the entire sky.
This homogeneity is further ensured by a significant overlap in the MK
standards used for each sample, and close scrutiny of the non-overlapping
standards to verify consistency.  The spectral types for the first sample of 
664 stars from the northern
hemisphere observed at DSO are recorded in Table 1. These spectral types are
multi-dimensional, as they include not only the temperature and luminosity
types, but also indices indicating abundance peculiarities and the degree
of chromospheric activity.

Chromospheric activity is evident in our spectra through emission reversals
in the cores of the \ion{Ca}{2} K \& H lines, and in more extreme situations,
infilling and emission in the hydrogen lines.  We have indicated these
different levels of chromospheric activity in the spectral types with the
following notation:  ``(k)'' indicates slight emission reversals or infilling
of the \ion{Ca}{2} K \& H lines are visible; ``k'' indicates emission reversals
are clearly evident in the \ion{Ca}{2} K \& H lines, but these emission
lines do not extend above the surrounding (pseudo) continuum; ``ke'' indicates
emission in the \ion{Ca}{2} K \& H lines above the surrounding (pseudo)
continuum, usually accompanied with infilling of the H$\beta$ line, and 
``kee'' indicates strong emission in \ion{Ca}{2} K \& H, H$\beta$ and perhaps 
even H$\gamma$ and H$\delta$.  Because chromospherically active stars tend
also to be variable, the chromospheric activity ``type'' will also vary.  
We have, therefore, noted the observation date in the
notes to Table 1 for those stars which have been designated either
``ke'' or ``kee''.  Stars of special astrophysical interest have been noted
in \S6 as well as in the notes to Table 1.  Specific dates of observation 
for all of the stars observed on this project can be found in 
the ``footers'' of the spectra themselves, available on the project website.

Spectral types are important to this project because 1) they provide the
first detailed look at our data and enable us to pick out peculiar and
astrophysically interesting stars and 2) accurate spectral types yield
beginning values for our determination of the basic physical parameters
and provide a check on the derived physical parameters.  In addition,
our spectral types enable us to refine the census of stars within
40pc of the sun.  Figure 1 shows an HR diagram based on the spectral types in
Table 1 and {\it Hipparcos} parallaxes \citep{esa97}.  Note the sharp lower 
edge to the main sequence and the good separation between the luminosity 
classes.  Note, 
however, the handful of stars scattering below the main sequence.  These stars,
without exception, have large parallax errors. Most of these stars are in
double and multiple systems, explaining the parallax errors.  Our spectral 
types confirm that these stars lie significantly beyond 40pc.  These stars 
are listed in Table 2.

\begin{deluxetable}{rrlrr}
\tablenum{2}
\tablewidth{0pt}
\tablecaption{Stars Beyond 40 parsecs identified by MK Classification}
\tablehead{\colhead{HIP} & \colhead{BD/HD} & \colhead{Spectral Type} & 
\colhead{$\pi \pm \sigma_{\pi}$ (mas)\tablenotemark{a}} & 
\colhead{d$_{\rm MK}$\tablenotemark{b}}}
\startdata
  1663 & 1651\phantom{C}    & kA7hA9mF0 III & $46.72 \pm 20.16$ & 290 \\
  1692 & 1690\phantom{C}    & K2 III        & $43.42 \pm 33.05$ & 340 \\
  7765 & 10182\phantom{C}    & K2 III        & $61.30 \pm 36.67$ & 320 \\
 24502 & 33959C   & F5 V          & $39.77 \pm 22.40$ & 95 \\
 30362 & 256294\phantom{C}   & B8 IVp        & $48.08 \pm 13.63$ & 1050 \\
 30756 & 257498\phantom{C}   & K0 IIIb       & $55.52 \pm 33.89$ & 230 \\
 35389 & +18 1563\phantom{C} &  A5 V        & $52.47 \pm 28.98$ & 370 \\
 76051 & +10 2868C & G2 V CH$-$0.3 & $44.59 \pm 20.41$ & 90 \\
 84581 & $-$07 4419B & A9 III      & $100.89 \pm 43.18$ & 275 \\
116869 & +13 5158B & G8 V+       & $44.81 \pm 60.22$  & 80 \\
117042 & 222788\phantom{C}    &  F3 V         &  $53.88 \pm 32.40$ & 170 \\
\enddata
\tablenotetext{a}{Hipparcos parallaxes and errors (ESA 1997).}
\tablenotetext{b}{Approximate distance in parsecs based on the
MK type.}
\end{deluxetable}

\section{Basic Physical Parameters}

An important goal of this project is to derive the basic physical parameters -
the effective temperature, the surface gravity and the overall metallicity -
for as many of our program stars as possible.

\subsection{Determination of the Basic Physical Parameters}

To determine these parameters, we use a technique similar to that devised 
by \citet{gray01} which fits the observed spectra and fluxes from
medium-band (Str\"omgren $uvby$) and broad-band (Johnson and 
Johnson-Cousins $VRI$) photometry, and, when available, IUE spectra, to 
synthetic spectra and fluxes.  The fit
is achieved by minimizing a $\chi^2$ statistic formed from the 
point-to-point squared differences between the synthetic and observed
spectra plus a similar sum over the squared differences between the
observed and synthetic fluxes.  The sum of the squared differences for the 
spectra are given approximately three times the weight of the sum of the 
squared differences over the fluxes in the final $\chi^2$.  The synthetic 
spectra and fluxes are
based on \citet{kurucz93} {\sc atlas9} stellar atmosphere models (calculated
{\it without} convective overshoot) and the synthetic spectra are computed 
with the spectral synthesis program 
{\sc spectrum}\footnote{www.phys.appstate.edu/spectrum/spectrum.html} 
\citep{gray94}.  Since the publication of the Gray, Graham \&
Hoyt paper, much effort has been put into improving the spectral line
list used by {\sc spectrum} including updating the oscillator strengths
with the latest critically evaluated values from the NIST 
website\footnote{physics.nist.gov}.  
Full details can be found on the {\sc spectrum} website.

The minimization of the $\chi^2$ statistic is carried out by the 
multidimensional downhill simplex algorithm {\tt amoeba} 
\citep{press92}.  We have 
modified the Gray, Graham \& Hoyt technique by introducing a graphical
front end ({\tt xfit16}) to this algorithm, which allows the user 
to find visually an approximate global solution which is then polished 
using the {\sc simplex} engine.  Any of the four basic physical
parameters ($T_{\rm eff}$, $\log g$, $\xi_t$ -- the microturbulent velocity --
and [M/H] -- the overall metal abundance) may be held fixed or allowed
to vary in the solution.  A further improvement on the Gray et al. technique
is the possibility to introduce observed fluxes not only from Str\"omgren
$uvby$ photometry, but from Johnson and Johnson-Cousins $VRI$ photometry
and from IUE spectra.  In addition, {\tt xfit16} can rotationally broaden
the synthetic spectra, and thus is capable of treating even high $v \sin i$
stars.  The graphical program {\tt xfit16} can access multiple 
libraries of synthetic spectra for each of the different datasets in the
project (for
instance, the two dispersions from DSO, spectra from CTIO, the
David Dunlap Observatory  and the Bok telescope of Steward Observatory).
The program {\tt xfit16} allows the user not only to verify that the
solution obtained is the global solution, but to judge visually the quality of
the solution and to decide if any of the input data are defective.  While
{\tt xfit16} allows the user to deredden the observed fluxes, the reddening
for all of our program stars was assumed to be zero.

For the B- and early A-type stars in our program, we utilized observed 
fluxes from
the Str\"omgren $u$, $b$ and $y$ bands and, when available, IUE spectra
obtained from the MAST IUE website\footnote{http://archive.stsci.edu/iue/}.  
The flux calibration used for the Str\"omgren photometry is that of 
\citet{gray98}. 
Because the metallic lines in these stars are quite weak and thus do not
have much leverage on the solution, the solution was first optimized 
with [M/H] held fixed at 0.00, and then [M/H] was adjusted manually
until good agreement was obtained visually with the metallic-line
strengths.  The solution was then re-optimized and [M/H] readjusted if
necessary.  For these stars, a microturbulent velocity of 
2 km s$^{-1}$ was assumed. The IUE spectra, when available, were valuable 
in determining [M/H] because line blanketing is much greater in the 
ultraviolet.    An example of a {\sc simplex}/{\tt xfit16} solution for an
early A-type star, HD~47105, is shown in Figure 2.

For the late A- , F- and early G-type stars IUE spectra were generally 
not available, and so the flux solution was constrained by fluxes 
from Str\"omgren $u$, $b$ and $y$  photometry.  For these stars, all four 
physical parameters were allowed to vary to derive the final solution. 

For the late G- and K-type dwarfs, a number of points had to be taken into
consideration to achieve good solutions.  First, for these cooler stars
the flux solution is not well constrained by observed fluxes from the
Str\"omgren bands, but it is necessary to include photometric fluxes from
the red and the near infrared in the form of Johnson and/or Johnson-Cousins
$VRI$ photometry.  We have used the absolute flux calibration for the
Cousins $R$ and $I$ bands of \citet{bessell90}.  To ensure uniformity,
when only Johnson $VRI$ photometry was available for a star, we used the
equations of \citet{fernie83} to transform Johnson $V-R$ and $V-I$ colors
to Johnson-Cousins $V-R$ and $V-I$ colors and then used Bessell's calibration.
Unfortunately, many of our late G- and K-type stars do not have either
Johnson or Johnson-Cousins $VRI$ photometry.  We have found, however, that
for {\it dwarf} stars the following equations are able to predict 
Johnson-Cousins $V-R$ and $V-I$ colors from Str\"omgren $b-y$ colors
in the range $0.4 < b-y < 0.8$ with an accuracy of 0.015 magnitude for 
$V-R$ and 0.03 magnitude for $V-I$.  

\begin{displaymath}
V-R = 1.30(b-y) - 0.17
\end{displaymath}
\begin{displaymath}
V-I = 2.32(b-y) - 0.25
\end{displaymath}

We can detect only a slight dependence on metallicity in these
relationships, well within the errors for stars with [M/H]$ > -0.7$.  By
happy circumstance, most of the very metal-weak stars in our sample have
$VRI$ photometry.
The Kurucz fluxes in the Str\"omgren $u$ band do not reproduce well the 
observed fluxes in the late G- and K-type stars, and so the $u$ band 
fluxes are not used in the {\sc simplex} solution for these stars.

It is unfortunate that neither the energy distributions of the late G- and
K-type dwarfs nor our classification spectra strongly constrain the
surface gravity for these stars.  In the B, A, F and even early G-type
stars the Balmer jump strongly constrains the surface gravity, but this
feature is too weak to serve this purpose in the late G and K-type stars.
The normal MK luminosity criteria in the K-type stars (the strength of
the CN band, and certain lines of ionized species such as \ion{Sr}{2}
$\lambda4077$ and \ion{Y}{2} $\lambda4376$ -- both used in ratio with
adjacent \ion{Fe}{1} lines) do not offer sufficient leverage in the
{\sc simplex} solutions to yield accurate ($\pm 0.10$~dex) values for
$\log g$ in the {\it dwarfs}.  We have found that allowing $\log g$ to
be a free parameter in the {\sc simplex} solution for the late G- and
K-type dwarfs often results in an unreliable (usually too low) value for 
that parameter in the final solution.  We have therefore constrained 
$\log g$ to the value implied by the {\it Hipparcos} parallax and the 
mass-luminosity relationship \citep{gorda98}.  We have further constrained
$\xi_t = 1.0$km~s$^{-1}$.  With these constraints, we generally obtain
good to excellent solutions for dwarfs of spectral type K3 and earlier.
An example of such a fit for a K3 dwarf can be seen in Figure 3.

Later than K3, however, the quality of the synthetic spectra and fluxes, 
especially in the spectral region used (3800 -- 5600\AA) begins to 
deteriorate for the dwarf stars.  It is clear that in this spectral region, 
for $T_{\rm eff} <
4500$K, a significant contributer to the continuous opacity is missing
in both {\sc atlas9} and {\sc spectrum}.  Synthetic fluxes normalized
and matched in the red to observed fluxes similarly normalized are too
high compared to the observed fluxes in the blue-violet, and synthetic line 
strengths in the blue-violet are too strong.  This persists even when
both synthetic and observed spectra are normalized at a common point
in the blue-violet.  This effect is seen even in the NextGen models
and synthetic spectra of \citet{hauschildt99}.  This missing opacity 
in the blue-violet region is reminiscent of a similar effect found in 
K-giant stars in
the violet by \citet{short94}.  Short \& Lester suggested this
missing violet continuous opacity might be supplied by photodissociation 
of MgH.  However, \citet{weck03} have recently calculated this continuous
opacity; it is about an order of magnitude too small to explain the
observed effect.  Other possible culprits include CaH and the O$^-$ ion, as
CaH has a dissociation energy similar to that of MgH and the ionization
energy of O$^-$ is about 2 eV.
This discrepancy between observed and synthetic spectra and fluxes
unfortunately prevents us from calculating basic physical parameters
for dwarf stars later than K3 from our spectra.   
 
For the late-G and K-type giants and subgiants it is likewise necessary 
to introduce
certain constraints to derive good solutions.  For most of the G and
K giants and subgiants  in our program, $V-K$ photometry exists, and so 
we have elected
to use the Infrared Flux Method (IRFM) of \citet{blackwell94} to
derive starting conditions and constraints for the {\sc simplex} solutions.  
For the few giants without $V-K$ photometry, we have found that the following
equations can be used to predict $V-K$ colors to sufficient accuracy
($\pm 0.06$ magnitude) for our purposes:
\begin{displaymath}
V-K = 3.80(b-y) - 0.02
\end{displaymath}
\begin{displaymath}
V-K = 1.39(B-V)^2 - 0.98(B-V) + 1.87
\end{displaymath}
valid for $0.55 < b-y < 0.90$ and $0.90 < B-V < 1.50$ respectively.
The IRFM method uses the following polynomial to predict the effective
temperature (for $V-K > 0.35$):

\begin{equation}
T_{\rm eff} = 8862 - 2583(V-K) + 353.1(V-K)^2
\end{equation}

The luminosity of the star may be determined using the {\it Hipparcos}
parallaxes with bolometric corrections \citep{flower96}, and the radius may
be determined from 
the Stefan-Boltzmann law.  We then determine the mass for the star using 
the evolutionary tracks of \citet{claret95} for Z = 0.01 (as many 
of the giants are slightly metal weak) and thence the surface gravity.
The normal procedure with the giants is then to constrain the $T_{\rm eff}$
to the IRFM value in the {\sc simplex} solution, constrain 
$\xi_t = 1.0$km~s$^{-1}$, begin with $\log g$ as calculated above,
and [M/H] is manually adjusted so that the line strengths in the
synthetic spectrum are approximately correct.  The {\sc simplex} algorithm 
is then allowed to polish the solution.  

The basic physical parameters for this first set of program stars may be
found in Table 1.

\subsection{Reliability of the Basic Physical Parameters}

A thorough discussion of the errors associated with the {\sc simplex} method
may be found in \citet{gray01}, but since a number of improvements
to both the spectral synthesis program {\sc spectrum} and to the {\sc simplex}
method have been made since the publication of that paper, a review of the
errors is warranted.

An excellent internal check on the precision of the {\sc simplex} effective
temperatures is afforded by a comparison with our spectral types.  Figure 4
shows a plot of $T_{\rm eff}$ versus the spectral-type running number
\citep{keenan84} for the dwarfs in our sample.  A polynomial fit yields a 
scatter of $\pm 115$K, part of which is attributable to errors in the 
spectral types and the width of the spectral boxes.  This polynomial 
yields the effective temperature/spectral
type calibration for dwarf stars in Table 3.  We conservatively estimate 
a random error in the {\sc simplex} effective temperatures of $\pm 80$K.  
For an external check, 
we may compare the {\sc simplex} effective temperatures with those derived 
from the IRFM method \citep{blackwell94}.  To do this, we have
selected those dwarfs in Table 1 which have $V-K > 0.35$ and have plotted
(Figure 5) $V-K$ against the {\sc simplex} effective temperatures.
The solid line in that figure shows the IRFM calibration (see equation 1).  As
can be seen, the agreement is excellent; the scatter around the IRFM 
calibration is $\pm 115$K and a close comparison suggests that the
{\sc simplex} temperatures are systematically only 30K hotter than the
IRFM temperatures.

\begin{deluxetable}{cr|cr|cr}
\tablenum{3}
\tablewidth{0pt}
\tablecaption{An Effective Temperature Calibration for Dwarf Stars}
\tablehead{\colhead{SpT} & \colhead{$T_{\rm eff}$} & 
\colhead{SpT} & \colhead{$T_{\rm eff}$} & \colhead{SpT} 
& \colhead{$T_{\rm eff}$}}
\startdata
A2 & 8800 & F3 & 6740 & G5 & 5580 \\
A3 & 8480 & F5 & 6530 & G8 & 5430 \\
A5 & 8150 & F7 & 6250 & G9 & 5350 \\
A7 & 7830 & F8 & 6170 & K0 & 5280 \\
A9 & 7380 & F9 & 6010 & K1 & 5110 \\
F0 & 7240 & G0 & 5860 & K2 & 4940 \\
F1 & 7100 & G1 & 5790 & K3 & 4750 \\
F2 & 6980 & G2 & 5720 & \nodata & \nodata \\
\enddata
\end{deluxetable}

It is of interest to compare the {\sc simplex} effective temperatures with
those in the \citet{cayrel01} catalog 
(hereinafter the [Fe/H] catalog) 
which contains determinations of $T_{\rm eff}$, $\log g$ and [Fe/H] from
the literature based on high-resolution spectroscopy.  This catalog 
contains only data derived from digital detectors (i.e., older photographic 
determinations have been removed from this version of the catalog), but 
the data are not otherwise critically assessed.  The [Fe/H] values in this
catalog are derived using a number of techniques, ranging from the curve of
growth method to spectral synthesis.

Figure 6 shows that
for the G and K-type stars in common there is good agreement between 
the {\sc simplex}
values of $T_{\rm eff}$ and those of the [Fe/H] catalog ($\sigma = 100$K,
with a negligible zero-point difference), but in the F-type stars there
is a systematic difference in the sense that the {\sc simplex} effective
temperatures are about 100K hotter than the literature values.  There is,
however, good agreement at G0 and for the early F-type stars.

This systematic difference in the F-type stars can be traced to a faulty
convective over-shoot algorithm included in the {\sc atlas9} stellar
atmosphere program \citep{kurucz93}.  This faulty algorithm introduces a
systematic error in the structure of the stellar atmosphere models of the
partially convective F-type star atmospheres \citep{smalley97}.  This
produces a distortion in the temperature scale of the F-type stars
which is unfortunately reflected in many of the determinations listed in
the [Fe/H] catalog.

We have recomputed the {\sc atlas9} models for the F-type stars with 
convective overshoot turned off using the implementation of {\sc atlas9}
by Michael Lemke
\footnote{http://www.sternwarte.uni-erlangen.de/ftp/michael/atlas-lemke.tgz}
and used these models in the spectral libraries employed by {\sc simplex}
(see \citet{gray01} for further details).

An important parameter for those who would use our data for the nearby
stars in exo-planet searches is the metallicity [M/H] of the
star.  There is some indication that planets are found preferentially around
stars with higher than solar metallicities \citep{gonzalez99}.  

Comparing the {\sc simplex} [M/H] values with mean [Fe/H] values from the
[Fe/H] catalog for the stars in common, we find small but significant 
zero-point differences (see
Table 4).  The zero-point differences for the dwarfs are generally small,
but the zero-point difference for the G and K giants is quite large and 
requires some discussion.  Many of the determinations in the [Fe/H] catalog
for G and K-giants were made with stellar atmosphere models calculated
with the MARCS code or its predecessors \citep{gustafsson75}, and this 
may possibly
indicate a systematic difference between the Kurucz models and the MARCS
models.  However, we expect that this systematic difference in the giants
has more to do with the fact that the abundance determinations in the
[Fe/H] catalog were largely carried out at wavelengths $> 5000$\AA, and
for the most part in the red, whereas
the {\sc simplex} solutions use blue-violet spectra (3800 -- 5600\AA).  
We have noted above discrepancies in line strengths in the dwarfs between 
observed and synthetic spectra in the blue-violet for $T_{\rm eff} < 4500$K, 
and we expect that we are seeing a similar phenomenon in the giants, 
although to a lesser
degree (note that many of our giants have $T_{\rm eff} < 4500$K).  Missing
continuous opacity in the blue-violet would lead to greater line strengths
in the models, and thus systematically negative values for [M/H].  

\begin{deluxetable}{lcc}
\tablenum{4}
\tablewidth{200pt}
\tablecaption{Zeropoint Differences:  
[Fe/H]$_{\rm catalog} - $[M/H]$_{\rm{\sc simplex}}$}
\tablehead{\colhead{Stellar Type} & \colhead{Resolution} & 
\colhead{Zeropoint}}
\startdata
F \& G dwarfs  &  1.8 \AA\ & 0.02 \\
G \& K dwarfs & 3.6 \AA\ & 0.05 \\
G \& K giants & 3.6 \AA\ & 0.23 \\
\enddata
\end{deluxetable}

Having applied the zero-point corrections in Table 4 to the {\sc simplex}
[M/H] values,
we find excellent agreement with the [Fe/H] catalog values (see Figure 8).
The [M/H] values in Table 1 have had these corrections
applied.
The random error in the comparison is only $\pm 0.09$~dex, hardly 
larger than the internal scatter in the [Fe/H] catalog ($\approx 0.08$~dex,
illustrated with error bars on one point in the figure).  This is an 
impressive result, considering that the [Fe/H] catalog values
were determined from high resolution spectra.  Such accurate and
homogeneous [M/H] values 
for a large sample of nearby stars will make possible a number of 
investigations, including examination of the hypothesis that planets are
found
preferentially around metal-rich stars.  We will consider this hypothesis
in paper II of this series.

\section{Chromospheric Emission}

All of the spectra obtained for this project include the \ion{Ca}{2} K \& H
lines and thus can be used to obtain measures of the chromospheric emission,
as emission from the chromosphere can be detected in the cores of these
very strong lines.  This is an important measurement,
as chromospheric emission can be an indication of the age of a star
and/or its binary status.  An age determination can be important for
exo-planet searches, as an indication of a young age for a star might
preclude observations using interferometric or coronographic techniques 
due to the complicating effects of zodiacal light from a remnant 
protoplanetary disk.

We measure the chromospheric emission in our program stars by calculating
relative fluxes in four wavelength bands (see Figure 9), two of which
are centered on the \ion{Ca}{2} K \& H lines.  The other two bands measure
fluxes in the ``continuum'' just shortwards and longwards of \ion{Ca}{2}
K \& H.  These bands are essentially identical to those used in the Mount
Wilson chromospheric activity survey program \citep{baliunas95} except
that the bands centered on \ion{Ca}{2} K \& H are wider (4 \AA) than those
used at Mount Wilson (1 \AA) because our spectra are of lower resolution.
Our instrumental chromospheric emission index is calculated, like the 
Mount Wilson index, with the following equation:
\begin{displaymath}
S = 5\frac{K+H}{C_1+C_2}
\end{displaymath}

This index is defined in such a way that it is insensitive to the local
slope of the continuum in the vicinity of the \ion{Ca}{2} K \& H lines.
Thus, we find that this index is identical (within a few thousandths)
whether we use flux-calibrated spectra or uncalibrated raw counts.
We determine this index by using a feature in the spectral classification
program {\tt xmk19} that allows the user to shift the spectrum in
wavelength to exactly align the $K$ and $H$ bands with the cores of
the \ion{Ca}{2} K \& H lines and which then carries out a numerical
integration over the four bands.  To ensure accuracy in the numerical 
integration over these bands, we resample the spectrum into 0.1 \AA\ bins.

To place this chromospheric index on the Mount Wilson scale, it is necessary
to use ``standard'' stars to derive a transformation equation.  Unfortunately,
all solar-type stars show some variability in the chromospheric index,
and those with higher values of $S$ are generally more variable.  What this
means is that unless one has observations taken near in time to Mount Wilson
observations, it is impossible to derive an exact transformation equation.
One, however, can derive a transformation of sufficient accuracy to 
characterize stars as active, inactive, etc. by selecting from the Mount
Wilson list \citep{baliunas95} those stars that do not show long-term
secular trends or irregular behavior as calibration stars.  The stars we
have selected for this purpose are listed in Table 5.  The resulting 
calibration for the 1.8~\AA\ resolution
spectra is shown in Figure 10.  To fit the trend in Figure 10, we have
employed a cubic fit given by the following equation:
\begin{displaymath}
S_{\rm MW} = -39.976S_{18}^3 + 43.335S_{18}^2 - 11.592S_{18} + 1.041
\end{displaymath}
for $S_{18} < 0.361$.  For $S_{18} \ge 0.361$ (the inflection point of the
above cubic), the curve is extended with a straight line, having the slope
of the cubic at the inflection point:
\begin{displaymath}
S_{\rm MW} = 4.067S_{18} - 0.845
\end{displaymath}
This straight line is also used to extrapolate to values of $S_{\rm MW}$ 
outside the range represented by the calibration stars.

We estimate the errors in this transformation
allow a determination of the Mount Wilson activity index, 
$S_{\rm MW}$, to an accuracy of 5\% with larger uncertainties (including
unknown systematic errors) for $S_{\rm MW} > 0.8$.  To ensure compatibility of
the derived $S_{\rm MW}$ values from the 3.6~\AA\ resolution spectra with the
1.8\AA\ spectra, we first transform the instrumental $S_{36}$ values onto
the instrumental $S_{18}$ system and then use the above calibration.  
The transformation
from the $S_{36}$ system to the $S_{18}$ system is determined by using
stars observed with both resolutions within a time frame of a few
months.  This transformation is linear, with a slope of unity and a 
zeropoint difference of 0.039.  The scatter in this transformation, an
indication of the precision with which we can measure $S_{\rm MW}$, is
$\pm 0.010$.  

\begin{deluxetable}{rc|rc}
\tablenum{5}
\tablewidth{0pt}
\tablecaption{Chromospheric Activity Calibration Stars}
\tablehead{\colhead{HD} & \colhead{$<S_{\rm MW}>$} & \colhead{HD} &
\colhead{$<S_{\rm MW}>$}}
\startdata
9562  & 0.136 & 131156A & 0.461 \\
16160 & 0.226 & 136202\phantom{A} & 0.140  \\
16673 & 0.215 & 141004\phantom{A} & 0.155  \\
18256 & 0.185 & 142373\phantom{A} & 0.147  \\
22049 & 0.496 & 143761\phantom{A} & 0.150  \\
26923 & 0.287 & 152391\phantom{A} & 0.393  \\
29645 & 0.140 & 154417\phantom{A} & 0.269  \\
33608 & 0.214 & 158614\phantom{A} & 0.158  \\
43587 & 0.156 & 159332\phantom{A} & 0.144  \\
45067 & 0.141 & 178428\phantom{A} & 0.154  \\
78366 & 0.248 & 182101\phantom{A} & 0.216  \\
81809 & 0.172 & 187013\phantom{A} & 0.154  \\
82885 & 0.284 & 188512\phantom{A} & 0.136  \\
89744 & 0.137 & 190360\phantom{A} & 0.146  \\
100180 & 0.165 & 194012\phantom{A} & 0.198 \\
100563 & 0.202 & 201091\phantom{A} & 0.658 \\
106516 & 0.208 & 201092\phantom{A} & 0.986 \\
114378 & 0.244 & 206860\phantom{A} & 0.330 \\
114710 & 0.201 & 212754\phantom{A} & 0.140 \\
115383 & 0.313 & 216385\phantom{A} & 0.142 \\
120136 & 0.191 & 217014\phantom{A} & 0.149 \\
126053 & 0.165 & \nodata & \nodata \\
\enddata
\end{deluxetable}

The $S_{\rm MW}$ index measures the flux in the core of the \ion{Ca}{2}
K \& H lines, but there are both photospheric and chromospheric contributions
to this flux.  The photospheric flux may be removed approximately following
the procedure of \citet{noyes84} who derive a quantity $R_{\rm HK}
\propto F_{\rm HK}/\sigma T_{\rm eff}^4$ where $F_{\rm HK}$ is the flux
per cm$^{-2}$ in the H and K bandpasses.  $R_{\rm HK}$ can be derived from
$S_{\rm MW}$ by modeling the variation in the continuum fluxes (in bands
C$_1$ and C$_2$) as a function of effective temperature (using $B-V$ as
a proxy).  $R_{\rm HK}$ must then be further corrected by subtracting
off the photospheric contribution in the cores of the H \& K lines.  The
logarithm of the final quantity, $R^\prime_{\rm HK}$ is then a useful measure 
of the chromospheric emission, essentially independent
of the effective temperature.  We have calculated $R^\prime_{\rm HK}$ for
all of our dwarf program stars later than F5 and earlier than M0 
with $B-V$ photometry.  While
the conversion of $S_{\rm MW}$ into $R^\prime_{\rm HK}$ becomes increasingly
uncertain for $B-V > 1.20$ (approximately spectral type K5), we have
carried this calculation out for stars as late as K8.  Both the $S_{\rm MW}$
and the $\log R^\prime_{\rm HK}$ indices are tablulated for our program
stars in Table 1.

We follow \citet{henry96} in employing $\log R^\prime_{\rm HK}$ to
classify stars into ``Very Inactive'', ``Inactive'', ``Active'' and
``Very Active'' categories (see Figure 11 and Table 1).  The distribution 
of stars
in Figure 11 is similar to that in Figure 7 of Henry et al., except that
their study had a bias toward G-dwarfs and did not go to as late a 
spectral type as our project.  

Figure 12 shows a histogram of the $\log R^\prime_{\rm HK}$ values
for the program stars with $0.5 < B-V < 0.9$.  This distribution may
be compared directly with Figure 8 of Henry et al.  It is clear that
the distribution is bimodal, with a peak near $\log R^\prime_{\rm HK} = 
-5.0$ and another peak near $-4.5$, very similar to that found by
Henry et al.  We have used the Levenberg-Marquardt method 
\citep{press92} to fit a double-Gaussian model to this distribution;
the result is shown in Figure 12 with the residuals from the fit
illustrated in the lower panel.  It is clear that this double 
Gaussian-model (peaks at $\log R^\prime_{\rm HK} = -4.996$ and $-4.536$, with
FWHM = 0.266 and 0.352 respectively -- again very similar to the results 
of Henry et al.) fits the distribution within the errors, with the 
possible exception of one bin, at $\log R^\prime_{\rm HK} = -5.1$.  This 
bin shows an excess (a 2.5$\sigma$ result) of stars.  An examination of 
Table 1 shows that out of the 36 stars in this bin, only 2 are metal-weak
([M/H]$ < -0.50$) and 3 are slightly evolved (either classified as
subgiants or having $\log g <= 4.00$).   The remaining stars in this bin
are apparently normal, near-solar metallicity, 
chromospherically-inactive dwarfs.  This result suggests that the 
excess stars in this
bin and in all the bins representing very inactive stars 
($\log R^\prime_{\rm HK} < -5.1$) may be due to dwarfs in a 
Maunder-minimum phase (see as well Henry et al. 1996, and Table 6 for a
list of possible solar-metallicity Maunder-minimum dwarfs).   However, as 
these excesses are only marginally significant statistically with the current 
sample, we will defer discussion of this point to later papers in this series.

\begin{deluxetable}{rrlrr}
\tablenum{6}
\tablewidth{0pt}
\tablecaption{Possible Solar-Metallicity Maunder-Minimum Dwarfs}
\tablehead{\colhead{HIP} & \colhead{BD/HD} & \colhead{Spectral Type} & 
\colhead{[M/H]} & \colhead{$\log R^\prime_{\rm HK}$}}
\startdata
9269 & 12051 & G9 V & 0.00 & -5.134 \\
35872 & 57901 & K3- V & 0.00 & -5.135 \\
39064 & 65430 & K0 V & -0.20 & -5.110 \\
42499 & 73667 & K2 V & -0.15 & -5.127 \\
67246 & 120066 & G0 V & 0.14 & -5.193 \\
70873 & 127334 & G5 V CH+0.3 & -0.03 & -5.118 \\
88348 & 164922 & G9 V & 0.00 & -5.141 \\
101345 & 195564 & G2 V & -0.09 & -5.196 \\
116085 & 221354 & K0 V & 0.00 & -5.149 \\
\enddata
\end{deluxetable}

Most of the stars in the ``very active'' category in Figure 11 are well-known 
variables of either the BY~Dra or RS~CVn types.  These stars are
named in the notes to Table 1 (\S6).  Two exceptions are
HIP~90035 = BD+01~3657 which shows strong emission in \ion{Ca}{2} K \& H,
but which is not a known variable and BD$-$03~3040B which shows strong
emission in \ion{Ca}{2} and the hydrogen lines.  See the notes in \S6 and 
Figure 13.

\section{Notes on Astrophysically Interesting Stars}

\noindent HIP~7585 = HD~9986:  Solar twin.  Notice how closely the 
{\sc simplex} basic physical parameters resemble those of the sun.  This 
star is also 
chromospherically inactive, and has a classification spectrum 
indistinguishable from that of the sun.  However, in a speckle survey of 
G-dwarfs \citep{horch02}, this star, while not resolved, was listed as a 
suspected non-single star.\\
HIP~16209 = LHS~173: This star is clearly metal-weak, but 
abundance-independent criteria place the spectral type of this star near
to K7.  At that spectral type the MgH band at $\lambda4780$ appears of
nearly normal strength.  This luminosity (gravity) sensitive feature
therefore suggests an unusually high gravity.  Classified as a subdwarf
K7 star by \citet{gizis97}.\\
HIP~53910 = HD~95418: Some sharp lines, \ion{Fe}{2} 4233 enhanced.  May be 
mild shell star.\\
BD$-$03~3040B: strong emission in \ion{Ca}{2} H \& K and the hydrogen lines.  
This star is not a known variable, but has been detected in the X-ray by ROSAT
\citep{mason95}.  Observed on Jan 31, 2001.  Double with HD~96064.  
HD~96064 is as well chromospherically active suggesting that this is a 
young binary system.  See Figure 13.\\
HIP~90035 = BD+01~3657: Very active chromospherically, but is not a known 
variable. Observed on Aug 8, 2000.  Detected in the extreme ultraviolet 
\citep{lampton97}.  A probable new BY Dra variable.  See Figure 13.\\

\section{Conclusions}

We have presented a database for 664 dwarf and giant stars earlier than
M0 within 40pc of
the sun that includes new, homogeneous spectral types, basic physical
parameters and measures of chromospheric activity.  Similar data on the
remaining 2935 solar-type nearby stars will be presented in subsequent
papers in this series.  The goals of this project are to characterize 
the stellar
population in the solar neighborhood and to provide data which will be
useful in the selection of targets for the Space Interferometry Mission
and the Terrestrial Planet Finder Mission.  As an example of how this
database could be helpful in selection of targets for discovering
terrestrial planets around solar-type stars, the
following table (Table 7) contains stars from the database that satisfy
the following criteria:
\begin{enumerate}
\item Solar-type dwarfs with spectral types from F8 -- G8.
\item Solar metallicity: [M/H]$ > -0.10$ (following the hypothesis
of \citet{gonzalez99} that planets are found preferentially around metal-rich
stars -- the validity of this hypothesis will be 
tested in paper II).
\item Chromospherically inactive or very inactive: chromospheric activity
can be an indicator of a young age; to find terrestrial planets with
extraterrestrial life, older stars should be selected.  In addition,
a young age for the star might preclude observation by a spacecraft like
the TPF because
of the presence of excessive zodiacal light in the system.
\item Single and non-variable.
\end{enumerate}

It should be noted that 13 of the stars in Table 1 already have known
planets.  These are HD~19994, 46375, 75732, 89744, 137759, 143761, 145675,
186427, 190360, 210277, 217014, 217107 and 222404.  These stars have not
been included in Table 7.

\begin{deluxetable}{rrl}
\tablenum{7}
\tablewidth{0pt}
\tablecaption{Candidate Stars for Terrestrial Planet Surveys}
\tablehead{\colhead{HIP} & \colhead{BD/HD} & \colhead{Spectral Type}}
\startdata
1499 & 1461 & G3 V \\    
7585 & 9986 & G2 V \\  
7918 & 10307 & G1 V \\  
14150 & 18803 & G6 V \\ 
14632 & 19373 & F9.5 V \\
14954 & 19994 & F8.5 V \\
24813 & 34411 & G1 V \\  
30545 & 45067 & F8 V \\  
41484 & 71148 & G1 V \\  
49081 & 86728 & G4 V \\  
49756 & 88072 & G3 V \\  
56242 & 100180 & F9.5 V \\
61053 & 108954 & F9 V \\ 
67246 & 120066 & G0 V \\ 
70873 & 127334 & G5 V CH+0.3 \\
73100 & 132254 & F8- V \\
74605 & 136064 & F8 V \\
79862 & 147044 & G0 V \\
85042 & 157347 & G3 V \\
85810 & 159222 & G1 V \\
89474 & 168009 & G1 V \\
90864 & 171067 & G6 V \\
94981 & 181655 & G5 V \\
96895 & 186408 & G1.5 V \\
100017 & 193664 & G0 V \\
101345 & 195564 & G2 V \\
102040 & 197076 & G1 V \\
103682 & 199960 & G1 V \\
\enddata
\end{deluxetable}

The data presented in this paper are currently available on the project's
website and work is continuing on the
remaining stars in the project which will be the subject of future papers
in this series.  As more results become available and the statistical
significance of our results improve, we intend to examine detailed features
of the distribution of chromospheric activity among solar-type stars, the
validity of the hypothesis that exoplanets are found preferentially
around metal-rich stars and to point out stars of astrophysical interest
and of interest to the SIM and TPF missions.

\acknowledgements

This work has been carried out under contract with NASA/JPL 
(JPL Contract 526270) and has been partially supported by grants from
the Vatican Observatory and Appalachian State University.  
This research made use of the SIMBAD database, operated at CDS, Strasbourg,
France.  We would also like to thank Jean-Claude Mermilliod for his
assistance in the compilation of photometry for our database.  Many
thanks to graduate student Kelly Kluttz, and to undergraduate
students Chris Jackolski, John Robertson, Corey Yost and Kate Hix, all
at Appalachian State University, for assistance in observing.

\begin{deluxetable}{rrlcrccrcrrcc}
\tabletypesize{\scriptsize}
\tablecaption{Spectral Types, Basic Physical Parameters and Chromospheric Indices}
\tablehead{\colhead{HIP} & \colhead{BD/HD} & \colhead{SpT} & \colhead{N1} & \colhead{$T_{\rm eff}$} & \colhead{$\log g$} & \colhead{$\xi_t$} & \colhead{$[M/H]$} & \colhead{N2} & \colhead{S$_{MW}$} & 
\colhead{$\log(R^\prime)$} & \colhead{AC} & \colhead{N3}} 
\startdata
   171 & 224930\phantom{B}   & G5 V Fe$-$1 & \nodata & 5502 & 4.27 & (1.0) & $-$0.64 & \nodata & 0.183 & $-$4.880 &  I & 1.8 \\
   400 & 225261\phantom{B}   & G9 V  & \nodata & 5250 & 4.40 & (1.0) & $-$0.43 & \nodata & 0.148 & $-$5.094 &  I & 3.6 \\
   518 & 123\phantom{B}      & G4 V  & * & 5718 & 4.50 & (1.0) &  0.06 & \nodata & 0.257 & $-$4.644 &  A & 1.8 \\
   544 & 166\phantom{B}      & G8 V  & \nodata & 5412 & 4.41 & (1.0) &  0.07 & \nodata & 0.379 & $-$4.458 &  A & 3.6 \\
   677 & 358\phantom{B}      & B8 IV-V Hg Mn & \nodata & 13098 & 3.91 & (2.0) & (0.0) & * & \nodata & \nodata & \nodata & \nodata \\ 
   746 & 432\phantom{B}      & F2 III  & \nodata & 6915 & 3.49 & 3.1 & $-$0.02  & \nodata & \nodata & \nodata & \nodata & \nodata \\
   974 & +26 8\phantom{B}    & K3 V  & \nodata & \nodata & \nodata & \nodata & \nodata & \nodata & 0.392 & $-$4.775 &  I & 3.6 \\
  1499 & 1461\phantom{B}     & G3 V  & \nodata & 5700 & 4.31 & (1.0) &  0.07 & \nodata & 0.151 & $-$5.059 &  I & 1.8 \\
\enddata
\tablecomments{The column headings have the following meanings: HIP stands for the desgination in the Hipparcos catalog \citet{esa97};
HD/BD for the HD or the BD designation; SpT for the spectral type -- note that some long spectral types are continued in the notes; 
N1: an asterisk in this column refers to a note on the spectral type; $T_{\rm eff}$, $\log g$, $\xi_t$, [M/H] refer to
the basic physical parameters -- the effective temperature (Kelvin), the logarithm of the surface gravity, the microturbulent velocity
in km~s$^{-1}$ and the overall metal abundance on a logarithmic scale, where [M/H] = 0.00 refers to solar metallicity (note that
if any of these quantities are enclosed in parentheses it was held fixed in the {\sc simplex} solution); N2: an asterisk in this
column refers to a note on the {\sc simplex} solution, a ``K'' in this column indicates that the method used to determine
the basic physical parameters was that adopted for the G \& K giants; S$_{MW}$ is the chromospheric activity index on the Mount 
Wilson system; $\log(R^\prime)$ is $\log R^\prime_{\rm HK}$, a measure of the chromospheric flux in the Ca~II K \& H lines; AC stands for the
chromospheric activity class: VI = Very Inactive, I = Inactive, A = Active, VA = Very Active; N3 indicates the resolution of the
spectrum in \AA ngstroms used for the measurement of S$_{MW}$ and $\log R^\prime_{\rm HK}$.  An asterisk in this column refers to a note on
chromospheric activity.}
\end{deluxetable}

Notes to Table1

In Table 1 there are three ``note'' columns pertaining to comments on the 
spectral types, the {\sc simplex} solutions and the measurements of
chromopheric activity which direct the reader to the comments on individual 
stars below.  In some of these comments, the rotational broadening
($v \sin i$) of the star is noted; these are visual estimates used in
the {\sc simplex} solutions and should not be taken as actual measurements of
the $v \sin i$.

\noindent HIP~518 = HD~123 = V640~Cas.\\
HIP~677 = HD~358: Far UV in the IUE spectra is very discrepant with respect 
to the model, probably due to excess metal blanketing in this chemically
peculiar star.  The {\sc simplex} 
solution was carried out without IUE spectra, and is suspect.\\
HIP~2762 = HD~3196 = BU~Cet.\\
HIP~3937: Chromospherically active M-dwarf, observed on Dec 1, 2000.\\
HIP~7585 = HD~9986:  Solar twin.  Notice how closely the {\sc simplex} basic
physical parameters resemble those of the sun.  This star is also 
chromospherically inactive, and has a classification spectrum 
indistinguishable from that of the sun.  However, in a speckle survey of 
G-dwarfs \citep{horch02}, this star, while not resolved, was listed as a 
suspected non-single star.\\
HIP~7762 = BD+14 253: Double with HD~10182.  The {\it Hipparcos} parallax 
places this star outside the 40 pc limit.\\
HIP~7765 = HD~10182: Large parallax error, beyond 40pc.  See Table 2.\\
BD+63~241:  Most probably an optical, not physical double with HD~10780.\\
HIP~8796 = HD~11443: Rotational broadening $ = 50$ km~s$^{-1}$.\\
HIP~10064 = HD~13161: Rotational broadening $ = 100$ km~s$^{-1}$.\\
HIP~12706 = HD~16970: Rotational broadening $ = 150$ km~s$^{-1}$.\\
HIP~12828 = HD~17094: Strong metallic-line spectrum; some metals = F1.\\
HIP~13976 = HD~18632 = BZ~Cet.\\
HIP~14576 = HD~19356 = Algol.\\
HIP~16209 = LHS~173: This star is clearly metal-weak, but 
abundance-independent criteria place the spectral type of this star near
to K7.  At that spectral type the MgH band at $\lambda4780$ appears of
nearly normal strength.  This luminosity (gravity) sensitive feature
therefore suggests an unusually high gravity.  Classified as a subdwarf
K7 star by \citet{gizis97}.\\
HIP~17666 = HD~23439B: \ion{Ca}{1} 4227 strong, 4500 -- 4800\AA\ region 
appears ``veiled''.\\
HIP~17695: Chromospherically active M-dwarf, observed on Jan 31, 2001.\\
HIP~17749/50 = HD~23189A/B = GL~153A/B/C:  This visual triple system has 
confusing nomenclature in the Simbad database.  The brightest component (A) 
is the K2 star, the star immediately adjacent and second in brightness, 
which we identify as B, is an M2 star.  The third and faintest of the 
triple, C (but which is identified in Simbad as BD+68 278A), is a K6 star.\\
HD~24916B: Chromospherically active M-dwarf, observed on Dec 1, 2000.\\
HIP~18915 = HD~25329: Metals are weak, but \ion{Ca}{1} 4227 and the G band are 
almost normal for K3.\\
HIP~19255 = HD~25893 = V491~Per.\\
HIP~19335 = HD~25998 = V582~Per.\\
HIP~19855 = HD~26913 = V891~Tau.\\
HIP~19859 = HD~26923 = V774~Tau.\\
HIP~21482 = HD~283750 = V833~Tau: Chromospherically active K3 subgiant, 
observed on Dec 1, 2000.\\
HIP~21818 = HD~29697 = V834~Tau: Chromospherically active K4 dwarf, observed
on Jan 31, 2001.\\
HIP~22845 = HD~31295: A3 Va kB9.5mB9.5 $\lambda$ Boo; 
rotational broadening $ = 100$ km~s$^{-1}$; IUE spectra used in the 
{\sc simplex} solution.\\
HIP~23875 = HD~33111: Rotational broadening $ = 180$ km~s$^{-1}$; 
IUE spectra used in the {\sc simplex} solution.\\
HIP~23941 = HD~33256: Spectral type may also be written F5.5 V Fe-0.7.\\
HIP~25278 = HD~35296 = V1119~Tau.\\
HIP~26335 = HD~245409: Chromospherically active M-dwarf, observed on 
Oct 28, 2000.\\
HIP~27913 = HD~39587 = $\chi^1$~Ori: G-band slightly weak \& 
chromospherically active.\\
HIP~28360 = HD~40183: A1 IV-Vp kA1mA1.5 (Sr).  Mild Ap star; 
rotational broadening $ = 100$ km~s$^{-1}$.\\
HIP~28954 = HD~41593 = V1386~Ori.\\
HIP~29525 = HD~42807 = V1357~Ori.\\
HIP~30362 = HD~256294: B8 IV kB9 Helium-weak.\\
HIP~30419 = HD~44769: Rotational broadening $ = 100$ km~s$^{-1}$.\\
HIP~30630 = HD~45088 = OU~Gem.\\
HIP~30756 = HD~257498: Large parallax error, beyond 40pc.  See Table 2.\\
HIP~30757 = HD~45352: Double with HIP~30756 = HD~257498.  This star has
a parallax of $5.67 \pm 2.22$, and thus, like its companion, is beyond 40pc.\\
HIP~31681 = HD~47105: IUE spectra used in the {\sc simplex} solution.  See
Figure 2.\\
HIP~32349 = HD~48915 = Sirius: Only an SWP IUE spectrum was used in the
{\sc simplex} solution as the LWP and LWR spectra were all defective or
had large gaps.  The {\sc simplex} fit is excellent, except for the K-line 
which is too strong in the model.  This is consistent with the Am 
nature of Sirius.\\  
HD~48682B = BD+43 1596.  Probable optical double.\\
HIP~35550 = HD~56986: Rotational broadening $ = 100$ km~s$^{-1}$.  Spectral
type may also be written F2 V Fe-0.5.\\
HIP~35643 = HD~56963: Spectral type may also be written F2 V Fe-0.5.\\
HIP~38541 = HD~64090: K0: V Fe-3 CH-1.5.\\
HIP~40375 = HD~68834: Noisy near \ion{Ca}{2} K \& H, so chromospheric 
activity measures uncertain.\\
HIP~41307 = HD~71155: Rotational broadening $ = 150$ km~s$^{-1}$.\\
HIP~44127 = HD~76644: Rotational broadening $ = 100$ km~s$^{-1}$; 
IUE spectra used in 
the {\sc simplex} solution.\\
HIP~45075 = HD~78362:  Anomalous luminosity effect.\\
HIP~45963 = HD~80715 = BF~Lyn. Chromospherically active K2.5 dwarf, observed
on Dec 6, 2000.\\
HIP~46843 = HD~82443 = DX~Leo.\\
HIP~46977 = HD~82210 = DK~UMa.\\
HIP~47080 = HD~82885 = SV~LMi.\\
HIP~49018 = HD~86590 = DH~Leo: {\sc simplex} solution poorly converged.\\
HIP~49593 = HD~87696: Rotational broadening $ = 120$ km~s$^{-1}$; 
IUE spectra used in 
the {\sc simplex} solution.\\
HIP~49669 = HD~87901: Rotational broadening $ = 250$ km~s$^{-1}$; 
IUE spectra used in  the {\sc simplex} solution.\\
HIP~51502 = HD~90089: Spectral type may also be written F4 V Fe$-0.5$.\\
HIP~53910 = HD~95418: Some sharp lines, \ion{Fe}{2} 4233 enhanced.  May be 
mild shell star. IUE spectra used in the {\sc simplex} solution.\\
BD$-$03~3040B: strong emission in \ion{Ca}{2} H \& K and the hydrogen lines.  
This star is not a known variable, but has been detected in the X-ray by ROSAT
\citep{mason95}.  Observed on Jan 31, 2001.  Double with HD~96064.  
HD~96064 is as well chromospherically active suggesting that this is a 
young binary system.  See Figure 13.\\
HIP~54872 = HD~97603: Rotational broadening $ = 150$ km~s$^{-1}$; 
IUE spectra used in the {\sc simplex} solution.\\
HIP~56035 = HD~99747: Spectral type may also be written F5 V Fe$-1.0$.\\
HIP~57632 = HD~102647: Rotational broadening $ = 120$ km~s$^{-1}$; 
IUE spectra used in 
the {\sc simplex} solution.\\
HIP~58001 = HD~103287: Rotational broadening $ = 150$ km~s$^{-1}$; 
only SWP IUE spectrum 
available for the {\sc simplex} fit.\\
HIP~59504 = HD~106112: Anomalous luminosity effect.\\
HIP~59744 = HD~106591: Rotational broadening $ = 180$ km~s$^{-1}$; 
IUE spectra used in 
the {\sc simplex} solution.\\
HIP~61941 = HD110379/80: $\gamma$ Vir AB.\\
HIP~61960 = HD~110411: A3 Va kB9.5mA0 $\lambda$ Boo; 
rotational broadening $ = 150$ km~s$^{-1}$; IUE spectra used in the 
{\sc simplex} solution.\\
HIP~62956 = HD~112185: Good fit except K-line in the model is stronger 
than in the star, consistent with the spectral type.\\
HIP~63076 = HD~112429: Rotational broadening $ = 100$ km~s$^{-1}$. Spectral
type may also be written F1 V(n) Fe$-0.8$.\\
HIP~63503 = HD~113139: Rotational broadening $ = 100$ km~s$^{-1}$.\\
HIP~65477 = HD~116842: Rotational broadening $ = 180$ km~s$^{-1}$; IUE 
spectra used in the {\sc simplex} solution.\\
HIP~66249 = HD~118098: Rotational broadening $ = 180$ km~s$^{-1}$.\\
HIP~66252 = HD~118100 = EQ~Vir: emission in \ion{Ca}{2}  K \& H, H$\beta$ and 
H$\gamma$ filled in, SED unusual, violet end depressed. 
Observed May 11, 2001.\\
HIP~67301 = HD~120315: Rotational broadening $ = 120$ km~s$^{-1}$.\\
HIP~69400 = HD~124752: Binary with BD+68 771B.\\
HIP~69673 = HD~124897: K0 III CH-1 CN-0.5.\\
HIP~69732 = HD~125162: A3 Va kB9mB9 $\lambda$ Boo; 
rotational broadening $ = 100$ km~s$^{-1}$; IUE spectra used in the 
{\sc simplex} solution.\\
HIP~71631 = HD~129333 = EK~Dra: G5 V Fe-0.7 CH-1 (k); this star is a 
young solar analog - it is chromospherically very active, and is a single, 
rapidly rotating G-type star with spot-induced variations (see 
\citet{dorren94} and references therein).  The rapid rotation, which is just
barely detectable with our 3.6\AA\ resolution spectra, may be responsible 
for the metal-weak classification.  Note, however, that the {\sc simplex} 
solution gives a nearly solar metallicity.\\
HIP~72848 = HD~131511 = DE~Boo.\\
HIP~74434 = BD+19 2939B: Double with HD~135101.\\
HIP~75695 = HD~137909: The LWP IUE spectrum shows very broad, strong 
absorption due to metal blanketing.  The fit was carried out without the 
IUE spectrum, and thus is suspect.\\
HIP~76051 = BD+10 2868C: Double with HD~138455; $\pi = 44.59 \pm 20.41$mas, 
whereas the parallax of HD~138455 is $\pi = 15.61 \pm 3.71$mas \citep{esa97}.\\
HIP~76267 = HD~139006: Rotational broadening $ = 150$ km~s$^{-1}$; the 
LWP IUE spectrum 
inconsistent with the two SWP spectra, and was not used in the fit.\\
HIP~77760 = HD~142373: G0 V Fe-0.8 CH-0.5.\\
HIP~79492 = HD~145958 \& BD+13~3091B: Both components of this visual
binary system yielded precisely the same instrumental value of the
chromospheric activity index $S_{36}$.  The two components as well are
nearly identical, translating into the same value for 
$\log R^\prime_{\rm HK}$.\\
HIP~79607 = HD~146361 = TZ~CrB: An RS CVn variable.  Many of the lines 
are veiled, probably by continuum emission; H$\gamma$ filled in with 
emission. Double with HD~146362.\\
HIP~79607 = HD~146362: Double with HD~146361.\\
BD+07 3125B: Double with HD~146413.\\
HIP~80008 = HD~147365: Rotational broadening $ = 70$ km~s$^{-1}$.\\
HIP~80644 = HD~148467: Noisy spectrum.\\
HIP~81300 = HD~149661 = V2133~Oph.\\
HIP~82588 = HD~152391 = V2292~Oph.\\
HIP~83601 = HD~154417 = V2213~Oph.\\
HIP~84379 = HD~156164: Rotational broadening $ = 250$ km~s$^{-1}$.\\
HIP~87108 = HD~161868: Rotational broadening $ = 200$ km~s$^{-1}$; IUE 
spectra used in 
the {\sc simplex} solution.\\
HIP~88601 = HD~165341 = 70~Oph = V2391~Oph: Visual binary.  Spectral type 
and the {\sc simplex} solution are for the primary.\\
HIP~90035 = BD+01~3657: Very active chromospherically, but is not a known 
variable. Observed on Aug 8, 2000.  Detected in the extreme ultraviolet 
\citep{lampton97}.  A probable new BY Dra variable.  See Figure 13.\\
HIP~91009 = HD~234677 = BY~Dra: Strong emission in \ion{Ca}{2} H \& K, 
H$\beta$ and H$\gamma$ both filled in with emission. Observed on 
Aug 31, 1999.\\
HIP~91262 = HD~172167 = $\alpha$ Lyr = Vega: IUE spectra used in the 
{\sc simplex} solution.\\
HIP~93747 = HD~177724: Rotational broadening $ = 300$ km~s$^{-1}$; 
IUE spectra used in the {\sc simplex} solution.\\
HIP~93805 = HD~177756: Metals and helium slightly weak; 
rotational broadening $ = 150$ km~s$^{-1}$; \ion{He}{1} lines strong in 
the model compared to the star, consistent with spectral type.\\
HIP~95575 = HD~183255: Slightly peculiar.\\
HIP~97649 = HD~187642: Rotational broadening $ = 200$ km~s$^{-1}$; 
IUE spectra used in the {\sc simplex} solution.\\
HD~335248: Optical double with HIP~102851 = HD~198550.  No parallax; from 
its spectral type and magnitude, it is likely beyond 40pc.\\
HIP~104659 = HD~201891: G5 V Fe-2.5 CH-1.5.\\
HIP~105199 = HD~203280: Rotational broadening $ = 200$ km~s$^{-1}$; 
IUE spectra used in the {\sc simplex} solution.\\
HIP~106231 = BD+22~4409 = LO~Peg: Observed on Aug 19, 2000.\\
HIP~106897 = HD~206043: Rotational broadening $ = 120$ km~s$^{-1}$.\\
HIP~107350 = HD~206860 = HN~Peg: flare star.\\
HIP~108706 = V374~Peg: A chromospherically active M-dwarf, observed
Aug 23, 2000.\\
HIP~112870 = HD~216259: Note to spectral type: may be slightly metal-weak.\\
HIP~113829 = HD~217813 = MT~Peg: G1 V CH-0.4 (k).\\
HIP~114189 = HD~218396: F0+ V kA5mA5 ($\lambda$ Boo); see 
Gray \& Kaye (2001).\\
HIP~114379 = HD~218738 = KZ~And.\\
HIP~115147 = HD~220140 = V368~Cep.\\
HIP~116584 = HD~222107 = $\lambda$~And.\\
HIP~116928 = HD~222603: IUE spectra used in the {\sc simplex} solution.

\begin{figure}
\figurenum{1}
\plotone{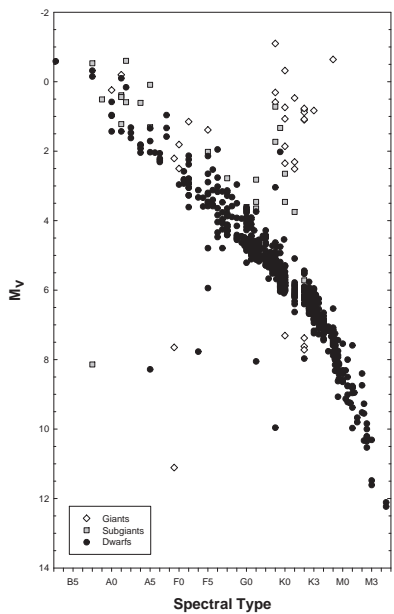}
\caption{An observational HR diagram formed from the spectral types of
the 664 stars reported in this paper.  The absolute magnitudes, $M_v$, are
calculated using {\it Hipparcos} (ESA 1997) parallaxes.  The stars
scattering below the main sequence in this figure all have large parallax
errors and are listed in Table 2.  These stars are all evidently more
distant than 40pc. }
\end{figure}

\begin{figure}
\figurenum{2}
\plotone{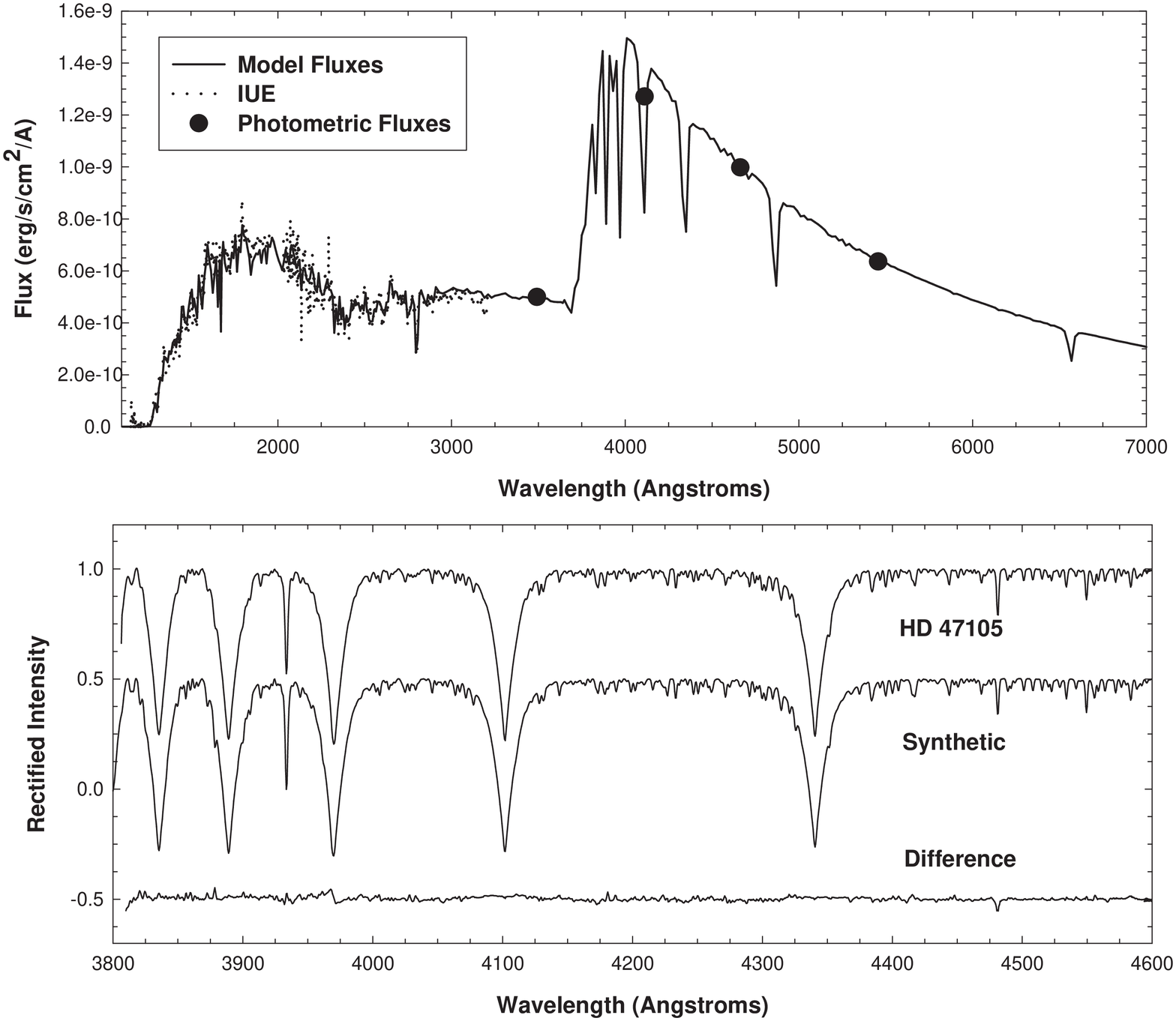}
\caption{The {\sc simplex} solution for the early A-type star HD~47105.  The
top panel shows the fit with respect to photometric fluxes (Str\"omgren
$uvby$) and ultraviolet fluxes from IUE spectra.  The bottom panel shows
the fit with the observed spectrum; the difference between the observed
and synthetic spectrum is seen at the bottom of the panel.}
\end{figure}

\begin{figure}
\figurenum{3}
\plotone{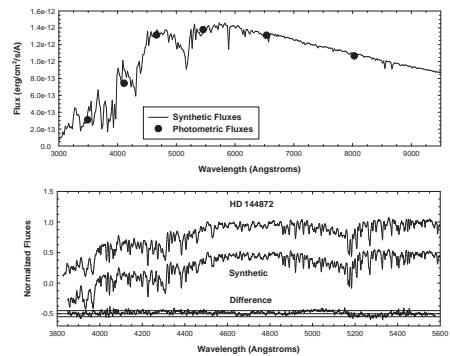}
\caption{An example of a {\sc simplex} fit for a K3 dwarf, HD~144872.  
The difference spectrum (synthetic minus observed) is shown at the
bottom of the second panel; the parallel lines indicate errors of
$\pm 5$\% in the flux.  The top panel shows the fit to the observed
photometric fluxes; the illustrated points are, from left to right,
Str\"omgren $u$, $v$, $b$ and $y$ fluxes and Johnson-Cousins
$R$ and $I$ fluxes.  The Str\"omgren $u$ flux is not used in the solution.}
\end{figure}

\begin{figure}
\figurenum{4}
\plotone{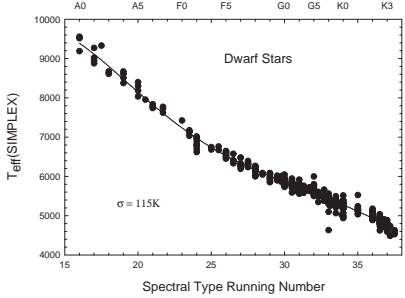}
\caption{The correlation between the {\sc simplex} effective temperatures
and the spectral type for the dwarf stars in the present sample.  The
scatter in this relationship is $\pm 115$K and sets an upper limit on the
random error in the {\sc simplex} effective temperatures.}
\end{figure}

\begin{figure}
\figurenum{5}
\plotone{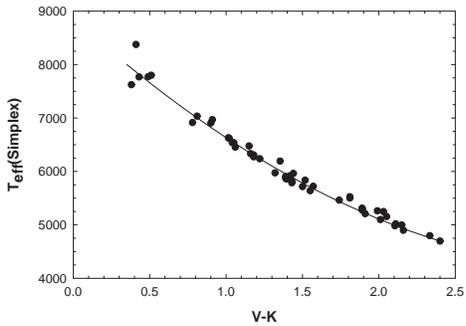}
\caption{Comparison of {\sc simplex} effective temperatures with the
IRFM calibration.  This comparison constitutes an external check on the
{\sc simplex} effective temperatures.  The solid line is the IRFM polynomial
(equation 1).  The scatter of {\sc simplex} temperatures around this
curve is 115K, and the systematic difference over this range of effective 
temperature does not appear to exceed 30K.  The IRFM temperatures are 
essentially independent of theoretical models.}
\end{figure}

\begin{figure}
\figurenum{6}
\plotone{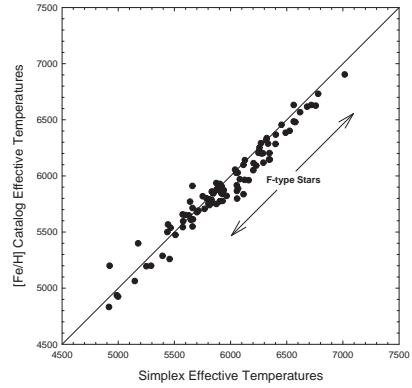}
\caption{A comparison of the {\sc simplex} effective temperatures with those
in the \citet{cayrel01} [Fe/H] catalog.  The systematic
difference in the F-type stars is probably due to a defective convective
overshoot algorithm used in the published {\sc atlas9} models (Kurucz 1993).
This has been corrected in the models used in the {\sc simplex} solutions.}
\end{figure}

\begin{figure}
\figurenum{7}
\plotone{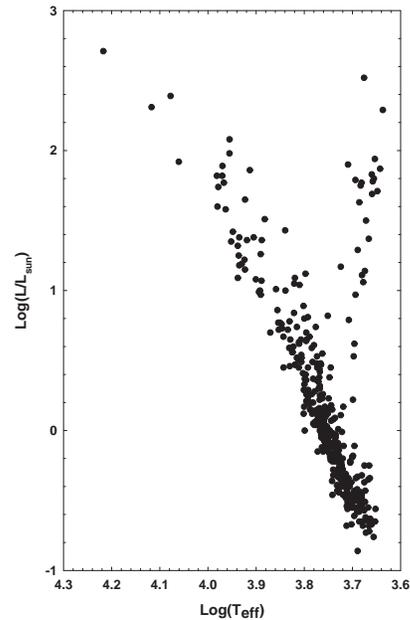}
\caption{An astrophysical HR diagram based on {\sc simplex} physical 
parameters.}
\end{figure}

\begin{figure}
\figurenum{8}
\plotone{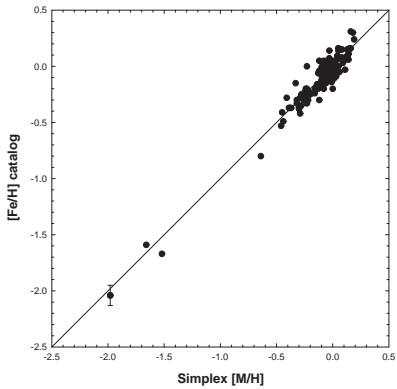}
\caption{A comparison between the {\sc simplex} [M/H] (overall metal
abundance) and mean [Fe/H] values from the \citet{cayrel01}
[Fe/H] catalog.  The scatter in the comparison is 0.09 dex, similar to
the scatter for single stars in the [Fe/H] catalog, $\approx 0.08$ dex.}
\end{figure}

\begin{figure}
\figurenum{9}
\plotone{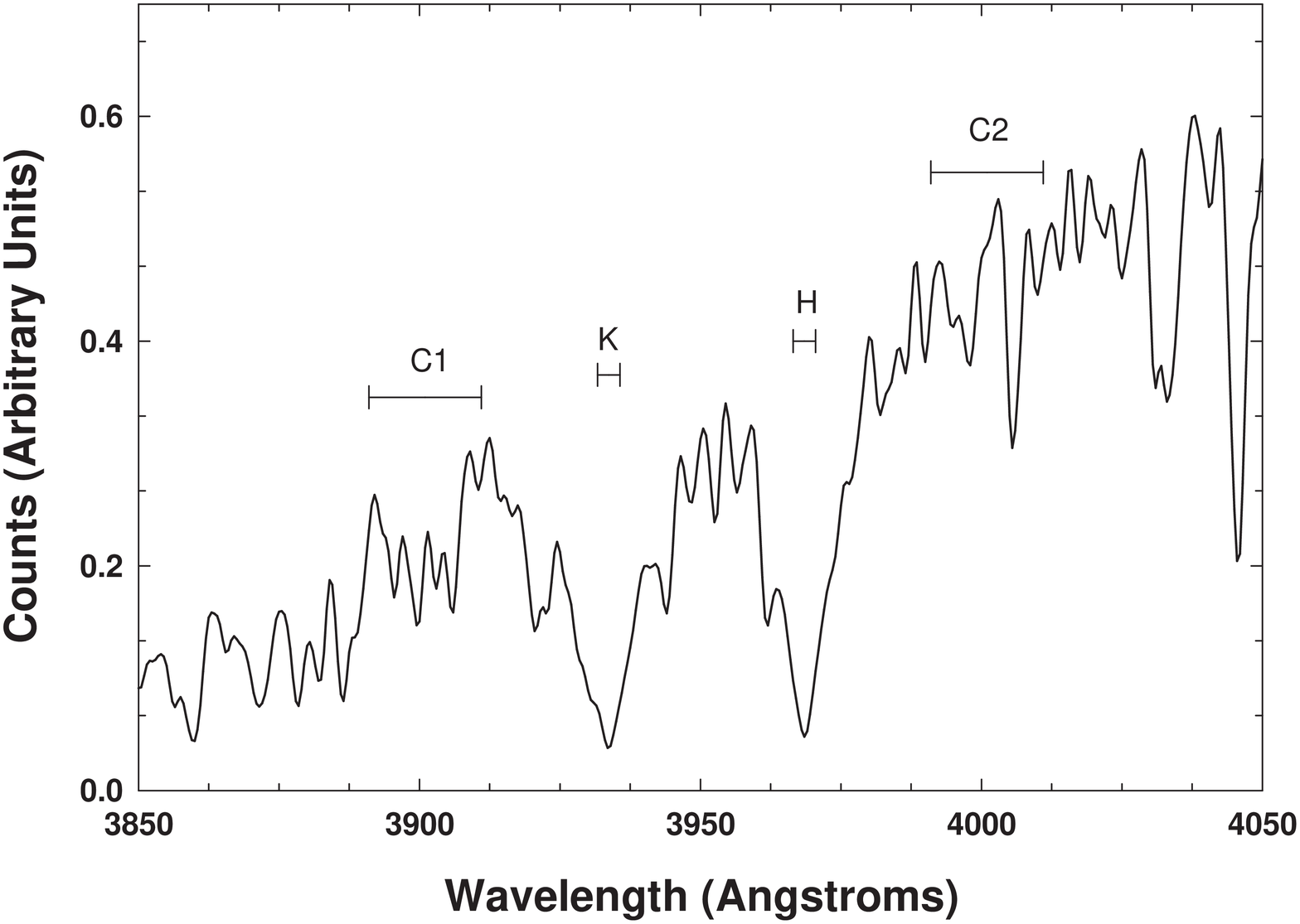}
\caption{Numerical passbands utilized in the calculation of the
chromospheric activity index $S$.  Two bands, each 4\AA\ wide, are centered
on the \ion{Ca}{2} K \& H lines.  The $C_1$ and $C_2$ bands measure the
flux in the continuum in 20\AA-wide segments of the spectrum on either side
of the K \& H lines.}
\end{figure}

\begin{figure}
\figurenum{10}
\plotone{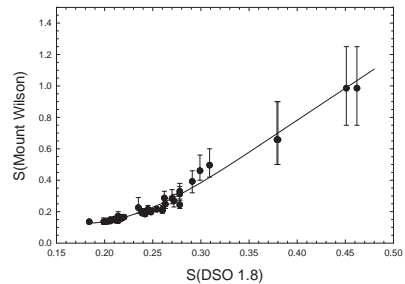}
\caption{The calibration between the instrumental $S_{18}$ chromospheric
activity index and the Mount Wilson $S$ index.  The ``error bars'' 
represent the range of variation shown by the star in Baliunas et al. (1995).
The solid line shows the adopted calibration.}
\end{figure}

\begin{figure}
\figurenum{11}
\plotone{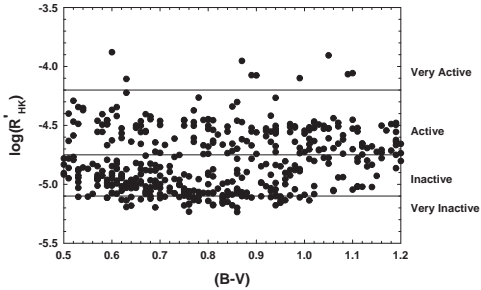}
\caption{A plot of the chromospheric flux parameter 
$\log R^{\prime}_{\rm HK}$ versus the $B-V$ color.  This diagram allows
the classification of stars into the chromospheric activity categories
``Very Inactive'', ``Inactive'', ``Active'' and ``Very Active''.  All
but one of the ``Very Active'' stars are well-known flare, BY Dra or RS CVn
variables.  The exception is HIP~90035 (see text and figure 13).}
\end{figure}

\begin{figure}
\figurenum{12}
\plotone{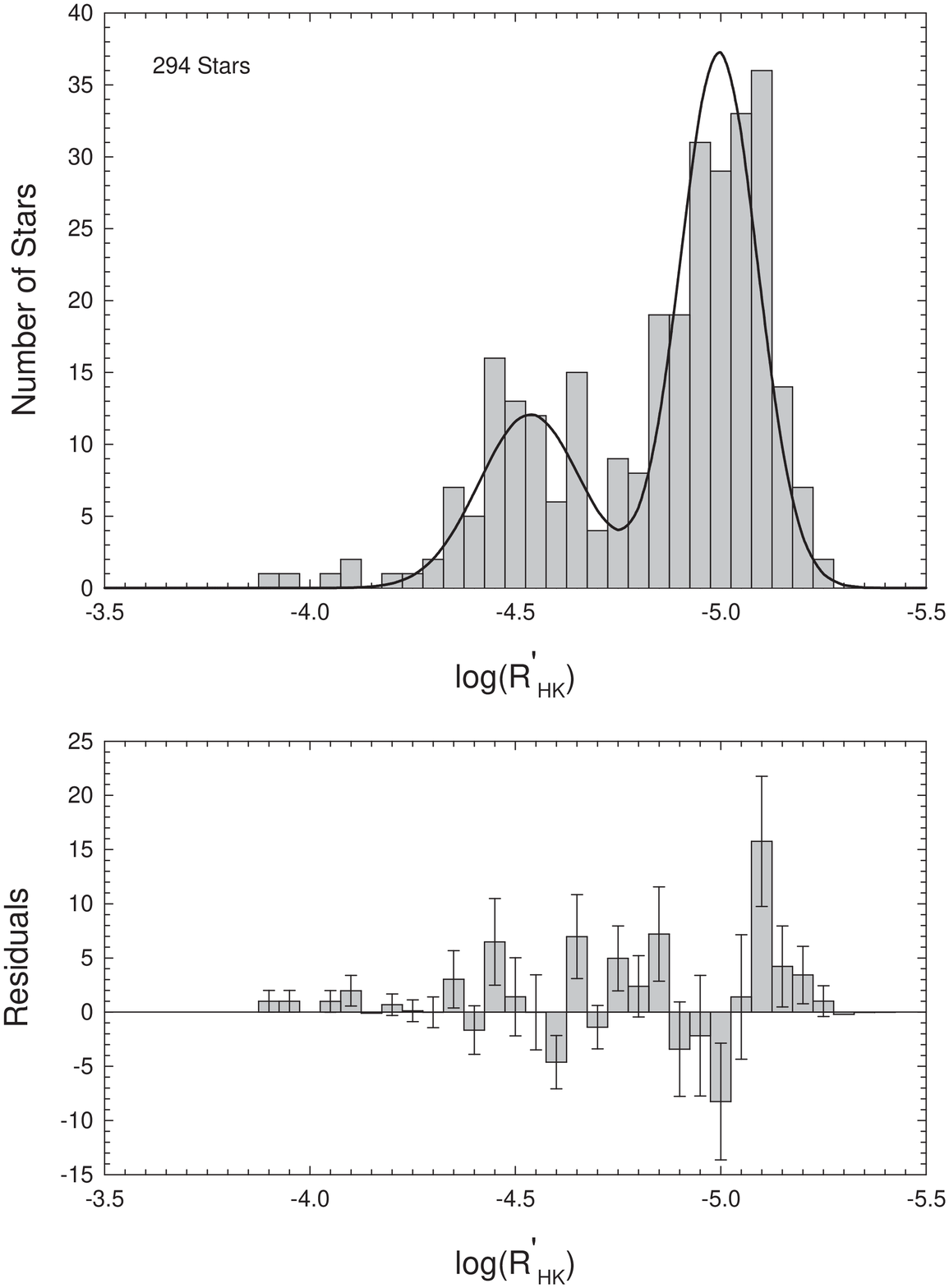}
\caption{The distribution of the chromospheric flux parameter
$\log R^{\prime}_{\rm HK}$ for stars with $0.50 < B-V < 0.90$.  The
distribution is clearly bimodal, with the main peak representing
chromospherically inactive stars, and the secondary peak chromospherically
active stars.  The double Gaussian curve is a fit to the distribution.
The lower panel shows the residuals associated with this fit.  With the
exception of the bin centered at $\log R^{\prime}_{\rm HK} = -5.1$,
the fit is within the errors.  The excess number of stars at
$\log R^{\prime}_{\rm HK} = -5.1$ is discussed in \S5. }
\end{figure}

\begin{figure}
\figurenum{13}
\plotone{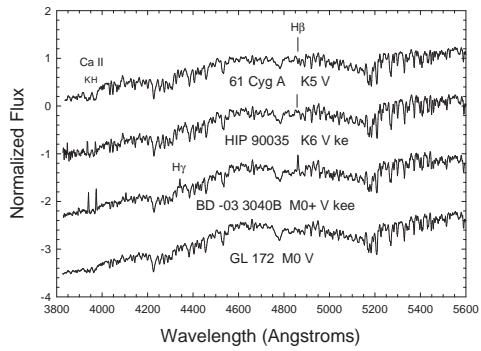}
\caption{HIP~90035 = BD+01~3657 is a chromospherically very active star
which has not been classified as a flare, BY Dra or RS CVn star.  Note
the strong emission (in comparison with the K5 V standard 61 Cyg A) in
\ion{Ca}{2} K \& H lines, and the infilling of the H$\beta$ line with
emission.  Likewise, BD~$-$03~3040B is a chromospherically active early-M
dwarf, not yet classified in the literature as an emission-line
star, showing strong emission in \ion{Ca}{2} K \& H as well as H$\beta$, 
H$\gamma$ and H$\delta$.}
\end{figure}

\end{document}